\documentclass[nopubldata]{lpor2014}
\usepackage{cite}
\usepackage{blindtext}
\usepackage{hyperref}
\hypersetup{%
  colorlinks,
  plainpages=false,
  pdfpagelabels,
  breaklinks,
  pdfview=Fit,
  bookmarksopen,
  bookmarksnumbered,
  linkcolor=black,
  anchorcolor=black,
  citecolor=black,
  filecolor=black,
  urlcolor=LPRred
  }%
\graphicspath{{./figs/}}
\setpages[]{}
\setvolume[0]{0}
\setyear{2011}%
\setdoi{201100000}%
\oddheadlogotext{}{Original Article}
\begin{document}

\titlefigure{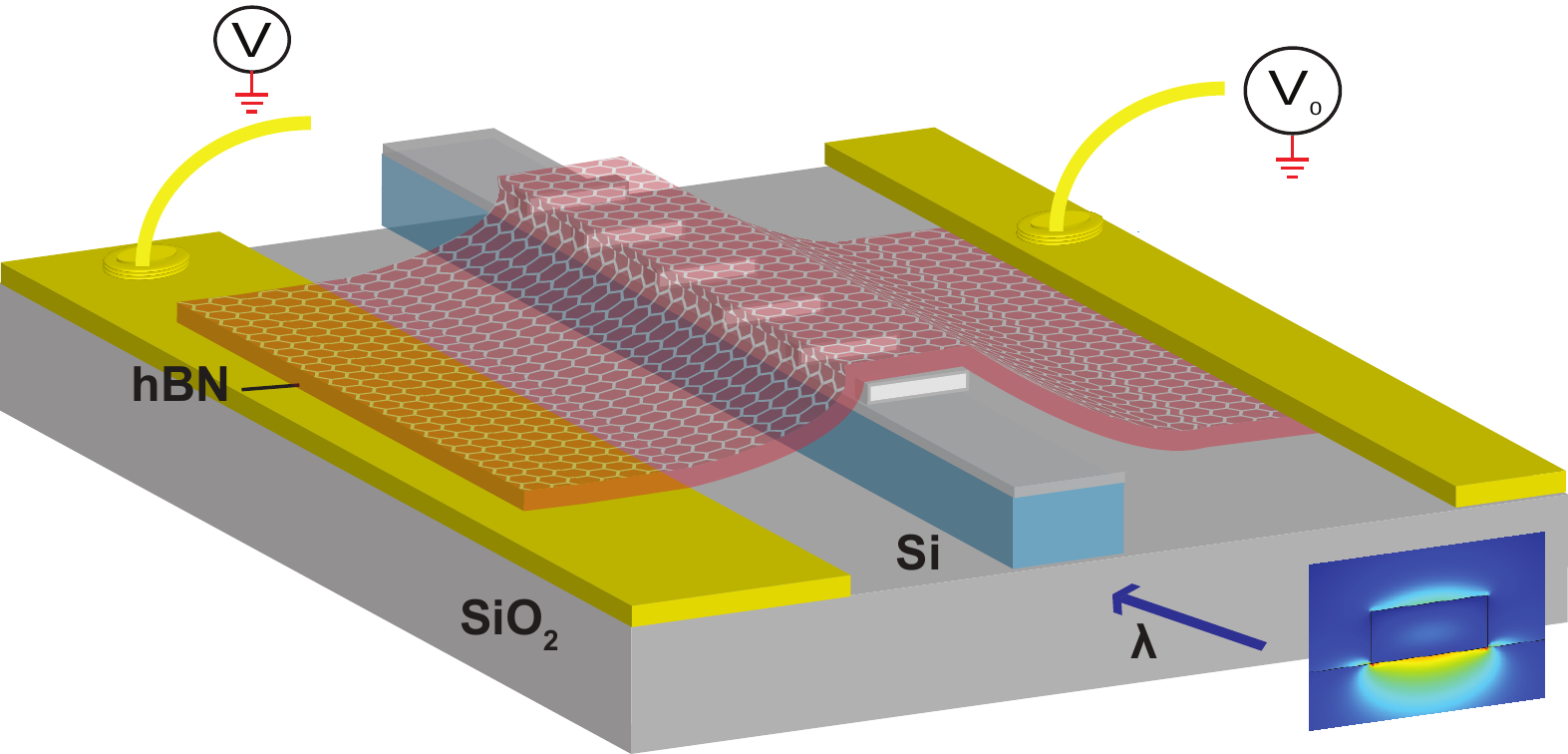}
\abstract{%
Graphene photonics has emerged as a promising platform for providing desirable optical functionality. However, graphene's monolayer-scale thickness fundamentally restricts the available light matter interaction, posing a critical design challenge for integrated devices, particularly in wavelength regimes where graphene plasmonics is untenable. While several plasmonic designs have been proposed to enhance graphene light interaction in these regimes, they suffer from substantial insertion loss due to metal absorption. Here we report a non-resonant metamaterial-based waveguide platform to overcome the design bottleneck associated with graphene device. Such metamaterial structure enables low insertion loss even though metal is being utilized. By examining waveguide dispersion characteristics via closed-form analysis, it is demonstrated that the metamaterial approach can provide optimized optical field that overlaps with the graphene monolayer. This enables graphene-based integrated components with superior optical performance. Specifically, the metamaterial-assisted graphene modulator can provide 5-fold improvement in extinction ratio compared to Si nanowire, while reducing insertion loss by one order magnitude compared to plasmonic structures. Such a waveguide configuration thus allows one to maximize the optical potential that graphene holds in the telecom and visible regimes.}

\title{Efficient integrated graphene photonics in the visible and near-IR}
\titlerunning{Efficient integrated graphene photonics}

\author{PoHan Chang\inst{1,*}, Charles Lin\inst{1} and Amr. S. Helmy\inst{1,*}}%
\authorrunning{PoHan Chang,}
\mail{\email{pohan.chang@mail.utoronto.ca, a.helmy@utoronto.ca}}

\institute{The Edward S. Rogers Department of Electrical and Computer Engineering, University of Toronto, 10 King's College Road, Toronto, Ontario M5S 3G4, Canada.}


\maketitle

\section{Introduction}
\label{sec:intro}
\paragraph*{}
The remarkable material properties of graphene have placed it as one of the most promising emerging photonic and optoelectronic materials.
The unique attributes such as ultra-wideband dispersionless nature \cite{Mak:2008}, ability to support high current density
\cite{Nair:2008,Bolotin:2008} and gate-variable optical conductivity \cite{Stauber:2008} have inspired a host of free-space as well integrated
optical components that incorporate graphene \cite{Ferrari:2010, Yu:2016, Goran:2016, Abdo:2016, Grigorenko:2012, Frank:2011, Daniel:2015, Ansell:2015, Radko:2016, Sorger:2016, Phatak:2016, Liu:2011, Sun:2016, Uros:2015}. These devices support a formidable range of functions, from
polarization manipulation, nonlinear conversion, plasmonic effects, light modulation and detection, to ultrafast components. Since the advent of
graphene photonics, other 2D materials with optical properties unattainable in bulk counterparts have also received profound
interest~\cite{Xia:2014}.
\begin{figure*}[t!]
\centering
\includegraphics[width=1.0\textwidth]{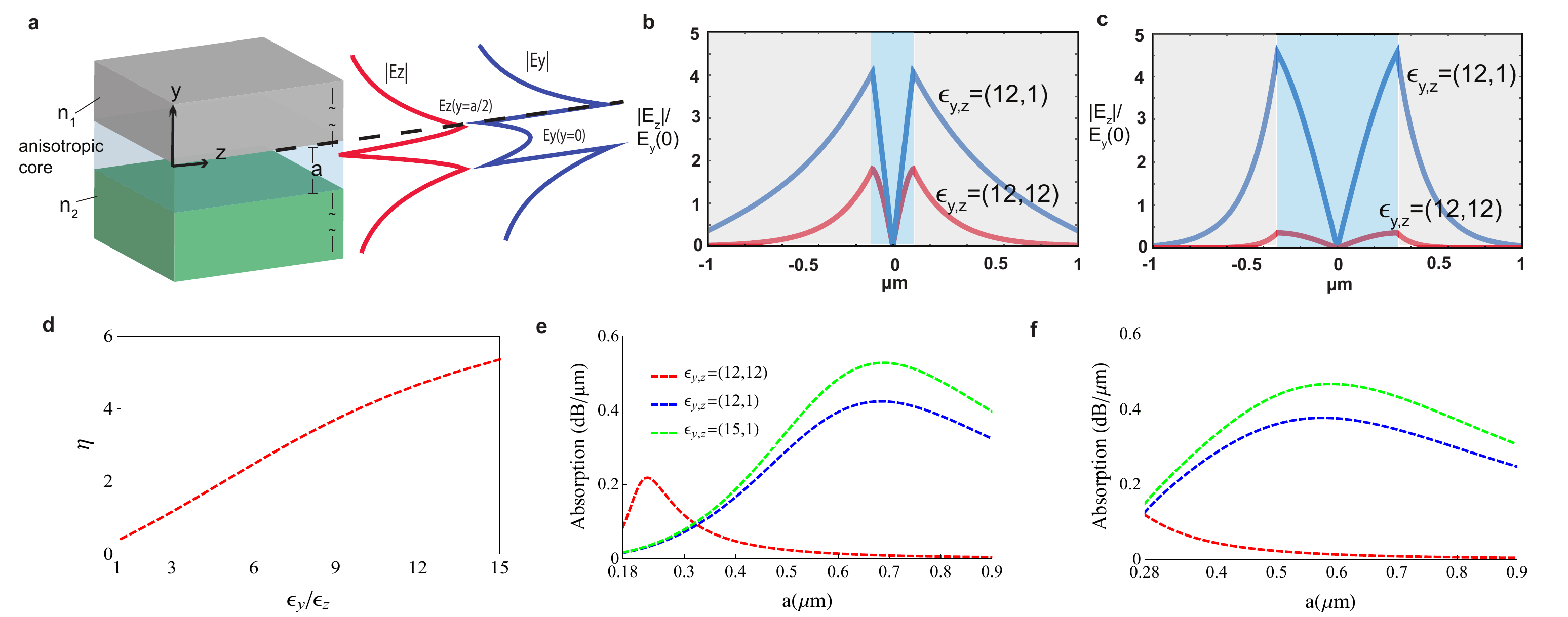}%
\caption{
\textbf{(a)} Schematic of a symmetric waveguide with high-index core. The field distributions of the 1D TM mode for an isotropic
waveguide are displayed ($\epsilon_{y}=\epsilon_{z}=12$). The dashed line indicates the placement position of graphene monolayer. \textbf{(b,c)}
Normalized $|E_{z}|$ distribution for isotropic and anisotropic waveguides with $a$ = 220 nm and 650 nm respectively. \textbf{(d)} Evolution of
$\eta$ in a 650 nm waveguide with varying degree of anisotropy ($\varepsilon_{y}=12$ is kept constant). \textbf{(e,f)} Absorption of the 1D TM
mode of symmetric ($n_{1,2}=1$) and asymmetric ($n_{1}=1, n_{2}=1.44$) graphene-cladded waveguides. The refractive index of graphene is taken to
be $n_{y}=1.6$ and $n_{z}=2.174-1.86i$ when the operational wavelength is 1550nm (See Suppl. Mat. for graphene model).
\label{Sym}}%
\end{figure*}

	The few-layer thickness associated with functional 2D films, in particular monolayer graphene, poses challenges to support efficient
light-matter interaction (LMI) between these layers and optical fields. In the mid-infrared (IR) regime, this may be circumvented by using
highly-confined graphene surface plasmon polaritons (SPP) modes~\cite{Grigorenko:2012}. However, within the visible and near IR frequency range
where graphene SPPs are not tenable, the limited LMI imposes significant performance bottleneck for optical components incorporating 2D
materials, particularly for guided-wave structures where the area of interaction is restricted. For example, graphene can be dynamically
reconfigured from absorptive to Pauli-blocked transparent state, allowing for high-speed, broadband modulators. However, modulator with graphene
overlaid on a Si-nanowire reported a suboptimal extinction ratio (ER) of only 0.15dB/$\mu$m~\cite{Liu:2011} due to the minimal overlap between
the optical mode and graphene sheet. While plasmonic or resonant waveguide structures can provide higher LMI and hence smaller form-factor, the
former suffer from substantial insertion loss (IL)~\cite{Ansell:2015, Radko:2016, Sorger:2016} while the latter have limited bandwidth and
temperature tolerance~\cite{Englund:2013,Qiu:2014}. To date, an athermal, low-loss guided-wave architecture that can better capitalize on graphene's optical attributes is still lacking.
	
	In this work, we report a metamaterial waveguide architecture that delivers a significant performance leap for integrated graphene photonics
operating within the visible and near-IR regimes, even though the light absorption is only attainable in the direction parallel to the graphene surface. More importantly, even though the metal is utilized here to enhance graphene LMI, the insertion loss of our metamaterial waveguide can be circumvented by one order of magnitude compared to existing designs using plasmonic structures. The novel metamaterial platform hence alleviates the loss-mode confinement tradeoff strictly dictated by graphene-based waveguide \cite{Charles:2015}. Using closed-form analysis, we demonstrate that extremely anisotropic metamaterial can enhance the in-plane field and modal confinement of guided wave structures simultaneously, thereby significantly improve the LMI obtainable within graphene-based devices. As illustrative examples, we examine the performance of graphene-based photodetectors and optical modulators consisting of hybrid metamaterial-dielectric waveguide core, which exhibit 2- and 5-fold enhancement in achievable absorption efficiency respectively. The waveguide can be implemented with hyperbolic~\cite{Alex:2013} or all-dielectric~\cite{Jacob:2016} metamaterial interlayer, and is compatible with
Si photonics~\cite{Neira:2014,Zayats:2015}. Such metamaterial-enabled platform can be extended to other 2D materials
\cite{Wang:2012,Charles:2016}, representing a new design paradigm for more compact, functional and integrated components.

\section{A guided wave solution for graphene-based waveguide}
To understand the conditions necessary to increase LMI between monolayer graphene and guided wave structures, we first examine the dispersion
properties of a 1D waveguide with high-index core and symmetric claddings (Fig.~\ref{Sym}(a)). Our analysis focuses on the z-propagating
transverse-magnetic (TM) mode, which is commonly utilized in integrated graphene photonics~\cite{Liu:2011,Ansell:2015, Radko:2016, Sorger:2016}.
The transfer matrix formalism \cite{Chen:2000} necessary for dispersion analysis is provided in the Supp Mat.
	
	Figure \ref{Sym}(a) inset depicts the electric field distribution of the 1D TM mode supported by a waveguide with isotropic core
($\epsilon_{y}=\epsilon_{z}=12$) of width $a$ = 220 nm. Since graphene's gate-tunable optical properties are only obtainable in the in-plane
directions, the $E_{y}$ component does not contribute to the overall LMI \cite{Kwon:2014}. Thus, to improve the performance of a graphene-cladded
optical device, the $E_{z}/E_{y}$ ratio for the optical mode should be maximized. For a symmetric waveguide, this ratio can be evaluated as (see
Suppl. Mat.):
	
\begin{equation}
\eta  = |E_{z,y=a}|/E_{y,y=a/2} = \frac{{{{\sin }}\left( {{k_y}a/2} \right){{\left( {{k_y}/{\varepsilon _z}}
\right)}}}}{{\frac{{{k_z}}}{{{\varepsilon _y}}}}} \propto \frac{\epsilon_{y}}{\epsilon_{z}}.\label{factor}
\end{equation}
However, eq. (1) reveals that an isotropic waveguide core ($\epsilon_{y}=\epsilon_{z}$) will always have limited LMI since
$\epsilon_{y}/\epsilon_{z}$ has a constant unity value. Such $\eta$ constraint can be resolved if the waveguide core exhibits extreme anisotropy
($\varepsilon_{z}\ll\varepsilon_{y}$). As evident in Fig.~\ref{Sym}(b), the $Ez$ component at the cladding-core interface can be enhanced by 2.3
times if the 220 nm isotropic core is replaced with an anisotropic one ($\epsilon_{cz}=1$ and $\epsilon_{cy}=12$).
	
\begin{figure*}[t!]
\centering
\includegraphics[width=0.9\textwidth]{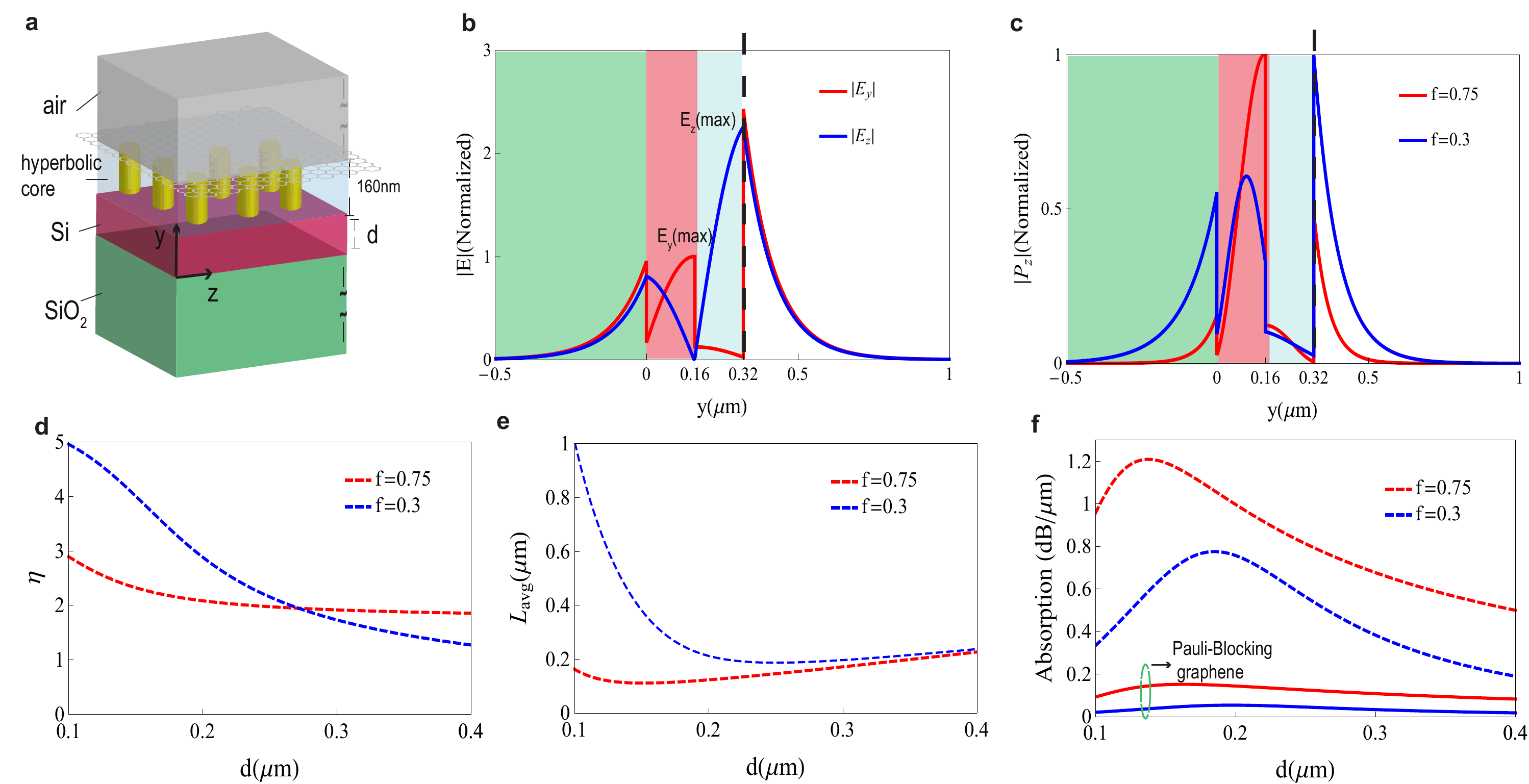}%
\caption{\textbf{(a)} Schematic of a graphene-cladded, hyperbolic metamaterial waveguide implemented using metal grating array. \textbf{(b)}
Normalized field distributions for hybrid waveguide with $d=160nm$ and $f=0.75$. \textbf{(c)} Normalized Poynting vector distributions of the
hybrid waveguide with varying volume filling fractions. \textbf{(d,e)} In-plane field enhancement ($\eta$) and modal confinement ($L_{avg}$) of
the hybrid hyperbolic waveguide as a function of the width of the dielectric core. \textbf{(e)} Absorption of the 1D TM mode of the hyperbolic
waveguide when graphene is in the absorptive state ($n_{y}=1.6$ and $n_{z}=2.174-1.86i$) and transparent state ($n_{y}=1.6$ and
$n_{z}=0.0027-1.416i$).
\label{Hyp}}%
\end{figure*}
	
	From Fig.~\ref{Sym}(b), it can be observed that the drawback with introducing anisotropy is that the optical modal confinement becomes
reduced since the optical momentum of the evanescent wave ($k_{y}$) decreases. Due to the slowly decaying evanescent field in the cladding
region, only $5\%$ of the modal power is contained in the anisotropic core region. The effective mode confinement ($L_{avg}$) can be quantified
by the ratio between the Poynting vector and the peak energy flux density (see Suppl. for discussion). Note that the effective mode confinement defined here refers to modal confinement for the entire core. Such lengthening of $L_{avg}$ can reduce the modal intensity overlapping graphene even though the relative in-plane field strength is enhanced via anisotropy. However, such trade-off between $\eta$ and  $L_{avg}$ can be alleviated by tuning the waveguide geometry (see Suppl. Mat. for discussion). Comparing Fig.~\ref{Sym}(b) and (c), it is observed that widening $a$ from 220 nm to 650 nm can significantly improve mode confinement, as $L_{avg}$ is be reduced from 3.5 to 2.8 $\mu m$  (Suppl. Mat. Fig. S1). Concurrently, the normalized $E_{z}$ at the waveguide interface is not only preserved, but is improved from 4.1 to 4.6.
	
	Figure.~\ref{Sym}(e) shows the absorption characteristics of the TM mode when graphene is overlaid onto the waveguide (indicated by the dashed line in Fig.~\ref{Sym}(a)), for both isotropic and anisotropic cores. For $a<350nm$,  isotropic waveguide outperforms its anisotropic counterpart due to weakened $L_{avg}$. However, light absorption of 0.4 $dB/\mu m$ can be obtained for wider anisotropic core ($a$ = 650 nm,
$\epsilon_{cz}=1$,  $\epsilon_{cy}=12$), clearly demonstrating the 2-fold performance enabled by the $\eta$-enhancement that is engineered via
anisotropy. The absorption efficiency can be further improved to 0.5 $dB/\mu m$ for higher anisotropy (Fig.~\ref{Sym}(d) and (e)).
	
	Note that the insights obtained from the symmetric waveguide can be extended to practical asymmetric waveguides ($n_{1}=1, n_{2}=1.44$) as
illustrated in Fig.~\ref{Sym}(f). In this case, utilization of anisotropic guiding layer can provide up to 4-fold improvement in absorption, an
upper limit ultimately dictated by the trade-off between $\eta$ and $L_{avg}$ factors. 
	
The trade-off between mode area and degree of anisotropy can be further optimized by implementing a hybrid metamaterial-dielectric waveguide
core. Such configuration is analogous to a hybrid plasmonic waveguide, which has dual cores where a top low-index core is utilized for light
confinement while a bottom high-index core is used to minimize propagation loss~\cite{Aitchison:2007,Oulton:2008,Wen:2014,Charles:2015}. In our
design, the metal layer is replaced with a top  metamaterial core to provide enhancement of in-plane field that is overlapping with graphene
while a bottom dielectric core is used to improve mode confinement. As shown in Fig. \ref{Hyp}(a), our analysis here will focus on hyperbolic
metamaterials that can be implemented using sub-wavelength silver gratings in air, but all-dielectric metamaterials are also applicable (see
Suppl. Mat.). Specifically, the anisotropic tensor of the grating can be homogeneously described via effective medium
theory~\cite{Agranovich:1985}:
\begin{equation}
{\varepsilon _\parallel } = {\varepsilon _x } = {\varepsilon _y } =  {f}{\varepsilon_{m}} + \left( {1 - {f}}
\right){\varepsilon_{0}},\label{par}
\end{equation}
\begin{equation}
{\varepsilon _ \bot } = {\varepsilon _z } =  \frac{{{\varepsilon_{m}}{\varepsilon_{0}}}}{{{f}{\varepsilon_{0}} + \left( {1 - {f}}
\right){\varepsilon_{m}}}}\label{per},
\end{equation}
where $f$ is the filling fraction of silver ($\epsilon_{m}=-130-3.3i$\cite{Johnson:1972}).
	
\begin{figure*}[t!]
\centering
\includegraphics[width=0.97\textwidth]{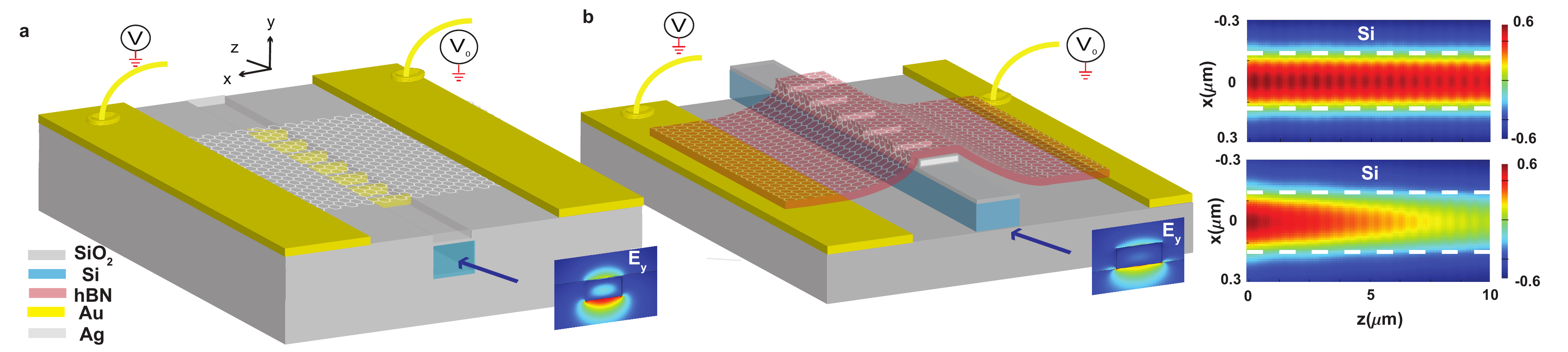}%
\caption{(a) Schematic of a hyperbolic metamaterial graphene photodetector. (b) Schematic of a hyperbolic metamaterial modulator design,
where the graphene is overlaid onto the grating structure. The insets show the field profiles of the incident TM modes. (c) Propagating electric
field profiles when graphene is in the transparent (top) and absorptive (bottom) states.
\label{Device}}%
\end{figure*}
	
	Figure~\ref{Hyp}(b) shows field profiles of the TM mode for a hybrid waveguide that consists of a bottom 160 nm Si core and a top 160 nm
hyperbolic core with $f$ = 0.75. Because $\varepsilon_{z}\ll\varepsilon_{y}$, Ez is enhanced while Ey becomes minimized within the hyperbolic
layer. The $f$ factor provides an additional degree of design freedom to tailor the field distribution (Fig.~\ref{Hyp}(c)). For example, as shown
in Fig.~\ref{Hyp}(d), hyperbolic core with smaller $f$ such as 0.3 has properties more sensitive to the dimension of the bottom Si core but can
provide stronger enhancement of $\eta$. Conversely, hyperbolic core with larger $f$ such as 0.75 allows for minimal $L_{avg}$ that is almost
invariant with core width (Fig.~\ref{Hyp}(e)). The light attenuation achievable by graphene-cladded hybrid metamaterial waveguide is displayed in
Fig.~\ref{Hyp}(f). The absorption efficiency when graphene is under Pauli-blocked transparent state are also plotted for comparison. Under
absorptive graphene state, the dual-core waveguide with $f$ = 0.75 allows for stronger attenuation of 1.1 dB/$\mu$m compared to 0.78 dB/$\mu$m
for $f$ = 0.3. In Pauli-blocked state, the design with $f$ = 0.3 allows for loss that is an order of magnitude lower than that of $f$ = 0.75. In
both cases, the hybrid dual-core waveguide offers drastic improvement in LMI over a single anisotropic core design. Thus, the $f$ factor should
be selected depending on the acceptable insertion loss. Such versatility, in turn, allows for a more complete LMI optimization specific to
different device applications. Note that graphene absorption may be further enhanced by the simultaneous optimization of both the anisotropic and high-index guiding layers to achieve optimal balance between $\eta$ and $L_{avg}$ of the waveguide structure.
\section{Graphene-based integrated devices}	
	To further elucidate the efficacy of metamaterial for improving LMI with graphene in the near-IR and visible ranges, here we study the
optical performance of simple integrated graphene photonic components that incorporate our hybrid architecture. First, a photodetector design
schematically shown in Fig.~\ref{Device}(a) is investigated. This is a common configuration where photocurrent is induced through light
absorption by a graphene monolayer that is directly overlaid on the guided wave structures~\cite{Andreas:2013,Gan:2013,Mo:2014,Wang:2016}. The
simulation set-up and waveguide parameters are described in Suppl. Mat. Based on 3D FDTD simulations, light attenuation at 1550 nm is calculated
to be 0.277 dB/$\mu$m for a 440 nm wide waveguide ($f$=0.6), which is an order of magnitude higher than previously reported near-IR graphene
detectors\cite{Andreas:2013,Gan:2013}. A typical metric for evaluating graphene LMI is the ratio of graphene absorption to the total device
absorption. Our simulations show that the waveguide is 0.02 dB/$\mu$m when graphene is removed, indicating a 90$\%$ graphene absorption
efficiency. Previous report of graphene modulator that also include metal layer only demonstrated 30-40$\%$ efficiency in the low-loss near-IR
regime~\cite{Andreas:2013}. In addition, our integrated metamaterial architecture can also operate in the visible regime where Ohmic loss will
strongly deteriorate device performance. We calculated an absorption efficiency of 33$\%$ at $\lambda=680nm$, the highest available to-date.
	
	Other than photodetectors, the strong LMI enabled by hybrid metamaterial-dielectric waveguide can also be utilized in optical modulators
(Fig.~\ref{Device}(b)). Based on 3D FDTD simulations, the ER and IL for a 300 nm wide device are calculated to be 0.5 dB/$\mu$m and 0.06
dB/$\mu$m respectively, as shown in Fig.~\ref{Device}(c). Table. \ref{Com} compares the simulated optical performance of various
graphene-assisted modulators. Our hybrid design exhibits comparable ER (0.5) to its plasmonic counterpart (1.2), showing a 5 times improvement
over that of Si nanowire. Nonetheless, our design does not suffer from significant IL as in the case of modulators that use plasmonic effect.
Such waveguide characteristics can also be supported in 1D structure (Fig.~\ref{Hyp}(f)). Because of $\varepsilon_{z}\ll\varepsilon_{y}$, the
dominant material absorption in our hyperbolic stack will occur in the y-direction ($Im(\varepsilon_{z})$$\ll$$Im(\varepsilon_{y})$) and should
increase with anisotropy. However, since anisotropy also leads to enhancement of $E_{z}$ and reduction of $E_{y}$ (Fig.~\ref{Hyp}(b)), the
overall material absorption from the hyperbolic layer will in fact decrease with increasing anisotropy. Such benefit cannot be obtained in
plasmonic-based designs because the isotropic dispersion of pure metal dictates that material absorption is identical in all directions, leading
to appreciable loss from the metal layer irrespective of relative intensity of $E_{z}$ and $E_{y}$.
	
\begin{table}
\center
\begin{tabular}{ |p{4.5cm}||p{1.75cm}|p{1.75cm}| }
\hline
Device Type& ER(dB/$\mu$m) &IL(dB/$\mu$m)\\
\hline
Si nanowire \cite{Liu:2011,Lum} & 0.106& 0.0112*\\
SPP waveguide \cite{Ansell:2015} & 0.002-0.2& ** \\
Dielectric-loaded plasmonic waveguide \cite{Radko:2016} &0.36 & 0.31\\
Slot plasmonic-waveguide \cite{Sorger:2016} &1.2 & 1.6\\
Our Metamaterial design&   0.5  & 0.06\\
\hline
\end{tabular}
\caption{Performance comparison of graphene-assisted modulators.* The IL originates from the absorption of graphene in transparent state
\cite{Lum} and scattering loss. ** The IL is the same order as ER \cite{Ansell:2015,Radko:2016}. \label{Com}}
\end{table}
\section{Conclusion}
In conclusion, we reported a metamaterial-based guided-wave platform that can significantly improve the performance of integrated graphene
photonic components. We showed that incorporating metamaterials in graphene photonics can provide enhanced graphene-guided wave overlap while
maintaining the necessary mode confinement. This leads to the realization of compact graphene-assisted devices, such as photodetectors and
modulators, with previously unattainable device performance. Moreover, we demonstrate that incorporation of hyperbolic effects is an effective
design strategy to alleviate the trade-off between ER and IL, allowing one to maximize the overall figure-of-merit of graphene modulators. This work allows, for the first time, the full modulation ability of graphene at telecom and visible wavelengths. More notably, such a low loss, non-resonant, and athermal architecture offers an optimal technology for designing next generation of integrated components utilizing 2D material.

%\newpage%%%%%%%%%%%%%%%%%%%%%%%%%%%%%%%%%%%%%%%%%%%%%
\bibliographystyle{lpr}

%Manual citation list

\end{document}